\documentstyle[12pt]{article} 
\input psfig.sty

\def\Title#1{\begin{center} {\Large #1 } \end{center}}
\def\Author#1{\begin{center}{ \sc #1} \end{center}}
\def\Address#1{\begin{center}{ \it #1} \end{center}}

\def\doeack{\footnote{Work supported by the Department of Energy,
                     contract DE--AC03--76SF00515.}}
\def\SLAC{Stanford Linear Accelerator Center\\
    Stanford University, Stanford, California 94309 USA}

\newenvironment{Abstract}{\begin{quotation} \begin{center}
                       ABSTRACT
     \end{center}\bigskip  }{\end{quotation}}
\def\beq{\begin{equation}}
\def\eeq#1{\label{#1}\end{equation}}
\def\eeqn{\end{equation}}
\def\beqa{\begin{eqnarray}}
\def\eeqa#1{\label{#1}\end{eqnarray}}
\def\eeqan{\end{eqnarray}}

\def\Re{{\cal R \mskip-4mu \lower.1ex \hbox{\it e}\,}}
\def\Im{{\cal I \mskip-5mu \lower.1ex \hbox{\it m}\,}}
\def\nn{\noindent}
\def\ie{{\it i.e.}}

\def\etal{{\it et al.}}

\def\sub#1{_{\lower.25ex\hbox{$\scriptstyle#1$}}}
\def\sul#1{_{\kern-.1em#1}}
\def\sll#1{_{\kern-.2em#1}}  
\def\sbl#1{_{\kern-.1em\lower.25ex\hbox{$\scriptstyle#1$}}}
\def\ssb#1{_{\lower.25ex\hbox{$\scriptscriptstyle#1$}}}
\def\sbb#1{_{\lower.4ex\hbox{$\scriptstyle#1$}}}

\def\to{\rightarrow}
\def\mh{\ifmmode m\sbl H \else $m\sbl H$\fi}
\def\mch{\ifmmode m_{H^\pm} \else $m_{H^\pm}$\fi}
\def\mt{\ifmmode m_t\else $m_t$\fi}
\def\mc{\ifmmode m_c\else $m_c$\fi}
\def\mz{\ifmmode M_Z\else $M_Z$\fi}
\def\mw{\ifmmode M_W\else $M_W$\fi}
\def\mws{\ifmmode M_W^2 \else $M_W^2$\fi}
\def\mhs{\ifmmode m_H^2 \else $m_H^2$\fi}   
\def\mzs{\ifmmode M_Z^2 \else $M_Z^2$\fi}
\def\mts{\ifmmode m_t^2 \else $m_t^2$\fi}
\def\mcs{\ifmmode m_c^2 \else $m_c^2$\fi}
\def\mchs{\ifmmode m_{H^\pm}^2 \else $m_{H^\pm}^2$\fi}
\def\ztwo{\ifmmode Z_2\else $Z_2$\fi}
\def\zone{\ifmmode Z_1\else $Z_1$\fi}
\def\mtwo{\ifmmode M_2\else $M_2$\fi}
\def\mone{\ifmmode M_1\else $M_1$\fi}
\def\tb{\ifmmode \tan\beta \else $\tan\beta$\fi}
\def\xw{\ifmmode x\sub w\else $x\sub w$\fi}
\def\ch{\ifmmode H^\pm \else $H^\pm$\fi}
\def\lum{\ifmmode {\cal L}\else ${\cal L}$\fi}
\def\inpb{\ifmmode {\rm pb}^{-1}\else ${\rm pb}^{-1}$\fi}
\def\infb{\ifmmode {\rm fb}^{-1}\else ${\rm fb}^{-1}$\fi}
\def\epem{\ifmmode e^+e^-\else $e^+e^-$\fi}
\def\ppb{\ifmmode \bar pp\else $\bar pp$\fi}

\def\bsg{\ifmmode b\rightarrow s\gamma \else $b\rightarrow s\gamma$\fi}
 
\newskip\zatskip \zatskip=0pt plus0pt minus0pt
\def\matth{\mathsurround=0pt}

\def\atversim#1#2{\lower0.7ex\vbox{\baselineskip\zatskip\lineskip\zatskip
  \lineskiplimit 0pt\ialign{$\matth#1\hfil##\hfil$\crcr#2\crcr\sim\crcr}}}

\begin{document}
\rightline{\vbox{\halign{&#\hfil\cr
&SLAC-PUB-7294\cr
&September 1996\cr}}}
\vspace{0.8in} 
\Title{Constraints on Anomalous Top Quark Couplings at the LHC}
\bigskip
\Author{Thomas G. Rizzo\doeack}
\Address{\SLAC}
\bigskip
\begin{Abstract}
Measurements of distributions associated with the pair production of top quarks 
at the LHC can be used to constrain (or observe) the anomalous chromomagnetic 
dipole moment($\kappa$) of the top. For example, using either the 
$t\bar t$ invariant mass or the $p_t$ distribution of top we find that 
sensitivities to $|\kappa|$ of order 0.05 are obtainable with 100 $fb^{-1}$ 
of integrated luminosity. This is similar in magnitude to what can be obtained 
at a 500 GeV NLC with an integrated luminosity of 50 $fb^{-1}$ through an 
examination of the $e^+e^- \to t\bar tg$ process.
\end{Abstract}
\bigskip
\vskip1.0in
\begin{center}
To appear in the {\it Proceedings of the 1996 DPF/DPB Summer Study on New
 Directions for High Energy Physics-Snowmass96}, Snowmass, CO, 
25 June-12 July, 1996. 
\end{center}
%
\bigskip
\def\thefootnote{\fnsymbol{footnote}}
\setcounter{footnote}{0}
\newpage

\section{Introduction}

The Standard Model(SM) has provided a remarkably successful description of 
almost all available data involving the strong and electroweak interactions. 
In particular, the discovery of the top quark at the Tevatron with a 
mass{\cite {tev}}, $m_t=175\pm 6$ GeV, close to that anticipated by fits to 
precision electroweak data{\cite {blondel}} is indeed a great triumph. 
However, we know that new physics beyond the SM must exist for many reasons 
particularly those associated with the fermion mass generating process. 
Since the top is the most massive 
fermion, it is believed by many that the detailed physics of the top quark 
may be significantly different than what is predicted by the SM. In this 
scenario, the top provides a window into the new physics which lies beyond the 
electroweak scale. This suggestion makes precision measurements of all of the 
top quark's properties absolutely mandatory and will require the existence of 
top quark factories. 

One of the most obvious and easily imagined scenarios is one in which 
the top's couplings to the SM gauge bosons, \ie, the $W$, $Z$, $\gamma$, 
and $g$, are altered. 
In the case of the electroweak interactions involved in 
top pair production in $e^+e^-$ collisions, the lowest dimensional 
gauge-invariant operators representing new physics that we can introduce take 
the form of dipole moment-type couplings to the $\gamma$ and $Z$. 
In the case of strong interactions, the subject of the present work,  
the corresponding lowest dimensional operator conserving $CP$
that we can introduce is the anomalous chromomagnetic moment
$\kappa${\cite {big,tgr}}. On the otherhand, the corresponding chromoelectric 
moment, $\tilde \kappa$, violates $CP$. 
In this modified version of QCD for the top quark the $t\bar tg$ interaction 
Lagrangian takes the form 
\begin{equation}
{\cal L}=g_s\bar t T_a \left( \gamma_\mu+{i\over {2m_t}}\sigma_{\mu\nu}
(\kappa-i\tilde \kappa \gamma_5)q^\nu\right)t G_a^\mu \,,
\end{equation}
where $g_s$ is the strong coupling constant, $m_t$ is the top quark mass, 
$T_a$ are the color generators, $G_a^\mu$ is the gluon field and 
$q$ is the outgoing gluon momentum. Due to the non-Abelian nature of 
QCD, a corresponding four-point $t\bar tgg$ interaction, proportional to 
$\kappa$ and/or $\tilde \kappa$, is by necessity also generated. 

Perhaps the most obvious place to probe for anomalous top couplings is at 
hadron colliders. It is clear that the existence of a non-zero value for 
$\kappa$ (and/or $\tilde \kappa$) would lead to a modification in {\it both} 
the $gg\to t\bar t$ and $q\bar q \to t\bar t$ subprocess cross sections at 
these machines. The 
general expressions for these parton level cross sections are given in 
Atwood \etal{\cite {big}}. Here we note only that the $q\bar q$ subprocess has 
a quadratic $\kappa$ dependence while that for the corresponding $gg$ 
subprocess has a quartic dependence on $\kappa$. In our discussion of 
anomalous top couplings at the 
LHC, we will ignore for brevity the possibility of a non-zero $\tilde \kappa$. 
Obviously, the observation of the $CP$-violation induced by non-zero 
$\tilde \kappa$ is a more 
sensitive probe for the anomalous chromoelectric moment of the top than the 
kinematic distributions we consider below.

\vspace*{-0.5cm}
\nn
\begin{figure}[htbp]
\centerline{
\psfig{figure=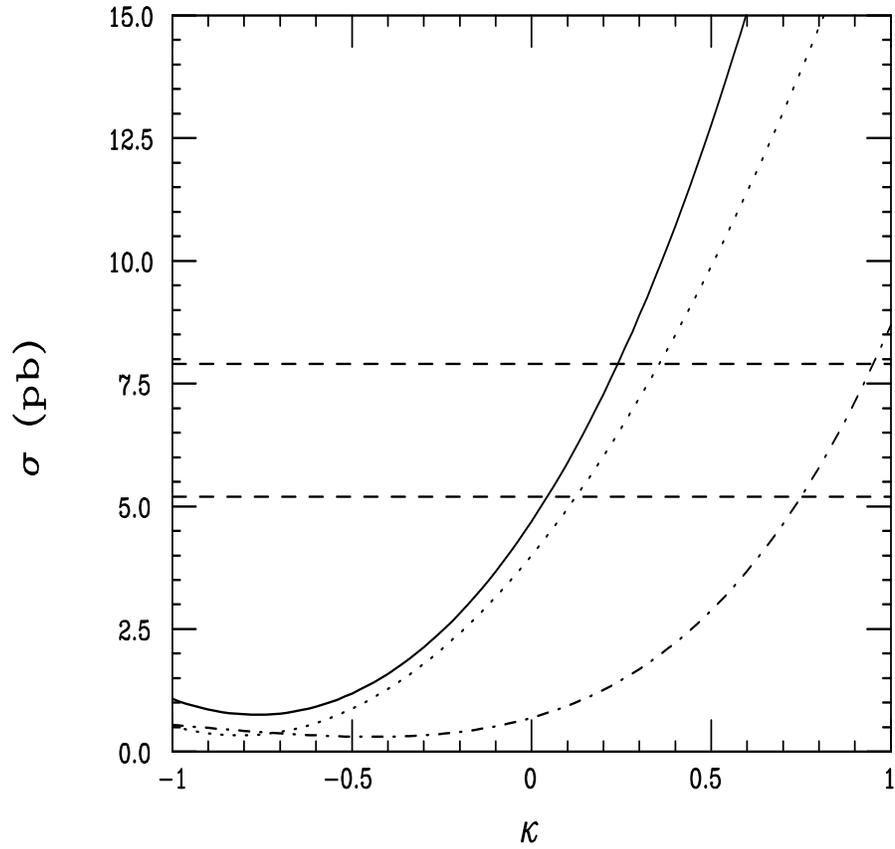,height=14cm,width=14cm,angle=-90}}
\vspace*{-1cm}
\caption{\small Cross section for $t\bar t$ production as a function of 
$\kappa$ at the Tevatron for 
$m_t=175$ GeV. The dotted(dash-dotted) curve is the $q\bar q(gg)$ contribution 
and the solid line is their sum. MRSA$'$ parton densities were assumed. The 
horizontal dashed bands correspond to the $1\sigma$ world average top pair 
cross section obtained by CDF and D0.}
\label{figtev}
\end{figure}
\vspace*{0.4mm}

\section{Effects of Anomalous Couplings}

At the Tevatron, it has been shown{\cite {big}} that for small values of 
$|\kappa| \leq 0.25$, a range consistent  with the current total cross 
section measurements{\cite {tev}} by both CDF and D0, the dominant effect of 
anomalous chromomagnetic moment couplings is to modify the total cross 
section for top pair production with little 
influence on the shape of the various distributions. Figure~\ref{figtev} 
compares the $\kappa$-dependent cross section with the world average of that 
obtained by the CDF and D0 Collaborations. 

The essential reason why the various top quark kinematical distributions are 
not much influenced 
is that top pair production at the Tevatron is dominated by the invariant 
mass region near threshold. Since, as is well known, the effects of 
anomalous couplings grow 
with the parton center of mass energy one sees little influence at these 
energies. The significantly larger partonic center of mass energies accessible  
at the LHC allows us to probe beyond this threshold region so that much higher 
sensitivities to a possible non-zero $\kappa$ can be obtained. This is 
particularly true for the various kinematic distributions. 

\vspace*{-0.5cm}
\nn
\begin{figure}[htbp]
\centerline{
\psfig{figure=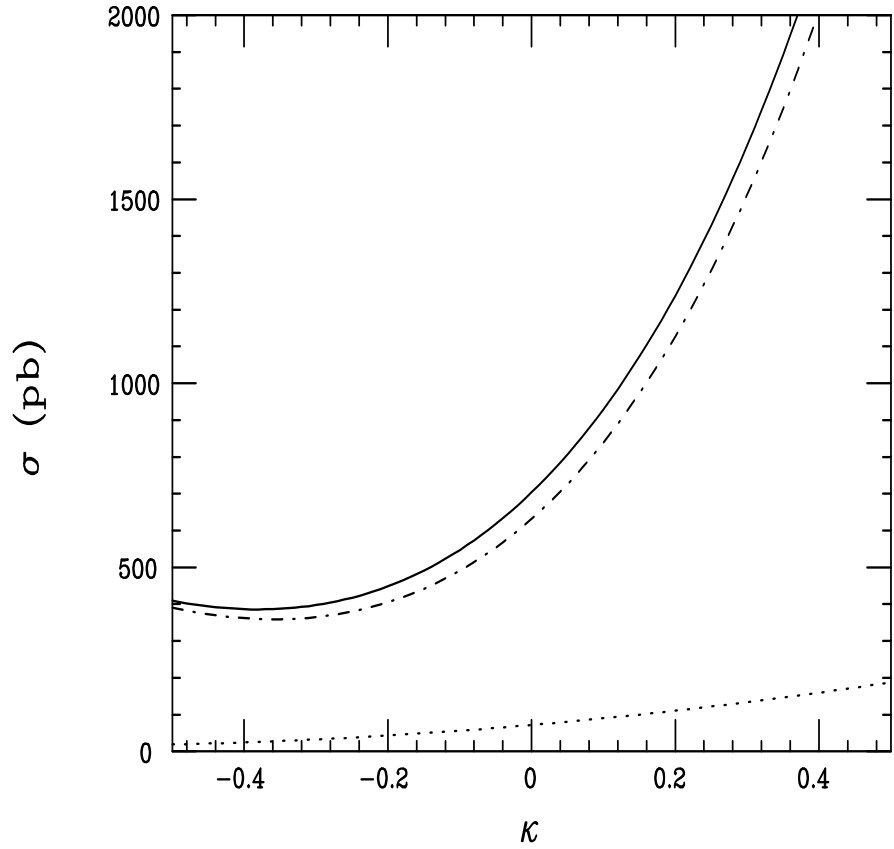,height=14cm,width=14cm,angle=-90}}
\vspace*{-1cm}
\caption{\small Cross section for $t\bar t$ production as a function of 
$\kappa$ at the LHC for 
$m_t=180$ GeV. The dotted(dash-dotted) curve is the $q\bar q(gg)$ contribution 
and the solid line is their sum. MRSA$'$ parton densities were assumed.}
\label{figlhc}
\end{figure}
\vspace*{0.4mm}

As a result of the subprocess dependencies on $\kappa$ it is 
clear than any of the 
(unnormalized!) differential distributions for a generic observable, $\cal O$,  
can then be written as 
\begin{equation}
{d\sigma \over {d\cal O}}= \sum_{n=0}^{4} \kappa^n g_n(\cal O)
\end{equation}
where $g_n(\cal O)$ are a set of calculable functions which have been 
completely determined to lowest order in QCD by Atwood \etal{\cite {big}}. 
The QCD/SM result is just the familiar term with $n=0$. Of 
course, the {\it total} cross section is also a quartic polynomial in 
$\kappa$. The behaviour of the two individual contributing subprocess as 
well as the total cross sections under 
variations of $\kappa$ at the LHC are shown in Fig.~\ref{figlhc}. Unlike the 
Tevatron, the $gg$ initial state dominates the top pair production cross 
section at the LHC. A reasonable sensitivity to $\kappa$ is again observed 
in the total cross section as it was for the Tevatron. However, 
as discussed in Ref.{\cite {big}}, unless the theoretical and systematic 
uncertainties are well under control, a measurement of $\sigma_{t\bar t}$ at 
the LHC will never do much better than to constrain $|\kappa|\leq 0.10-0.15$. 
To further improve on this limit we must turn to the various top quark 
kinematical distributions. 

\vspace*{-0.5cm}
\nn
\begin{figure}[htbp]
\centerline{
\psfig{figure=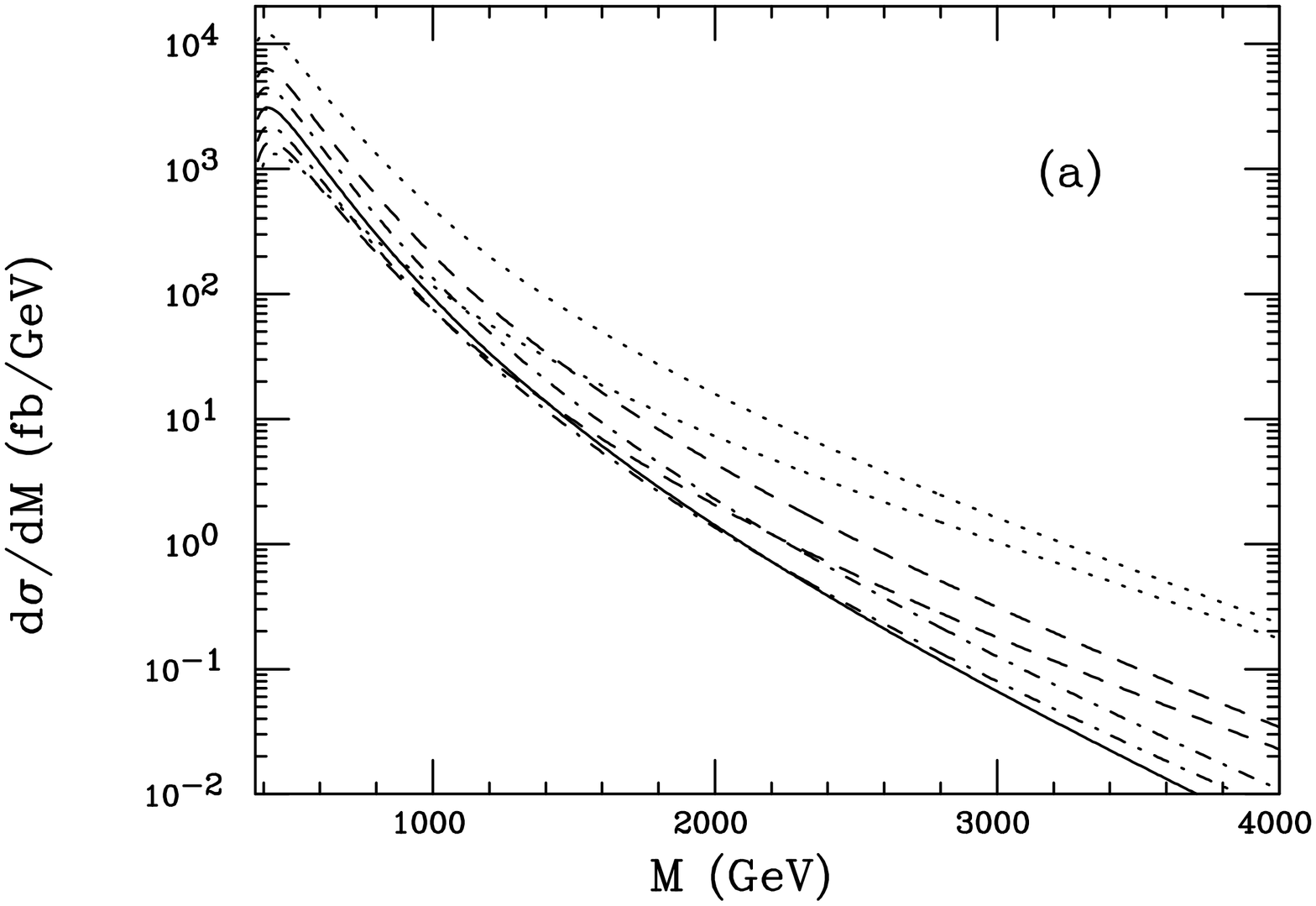,height=9.1cm,width=9.1cm,angle=-90}
\hspace*{-5mm}
\psfig{figure=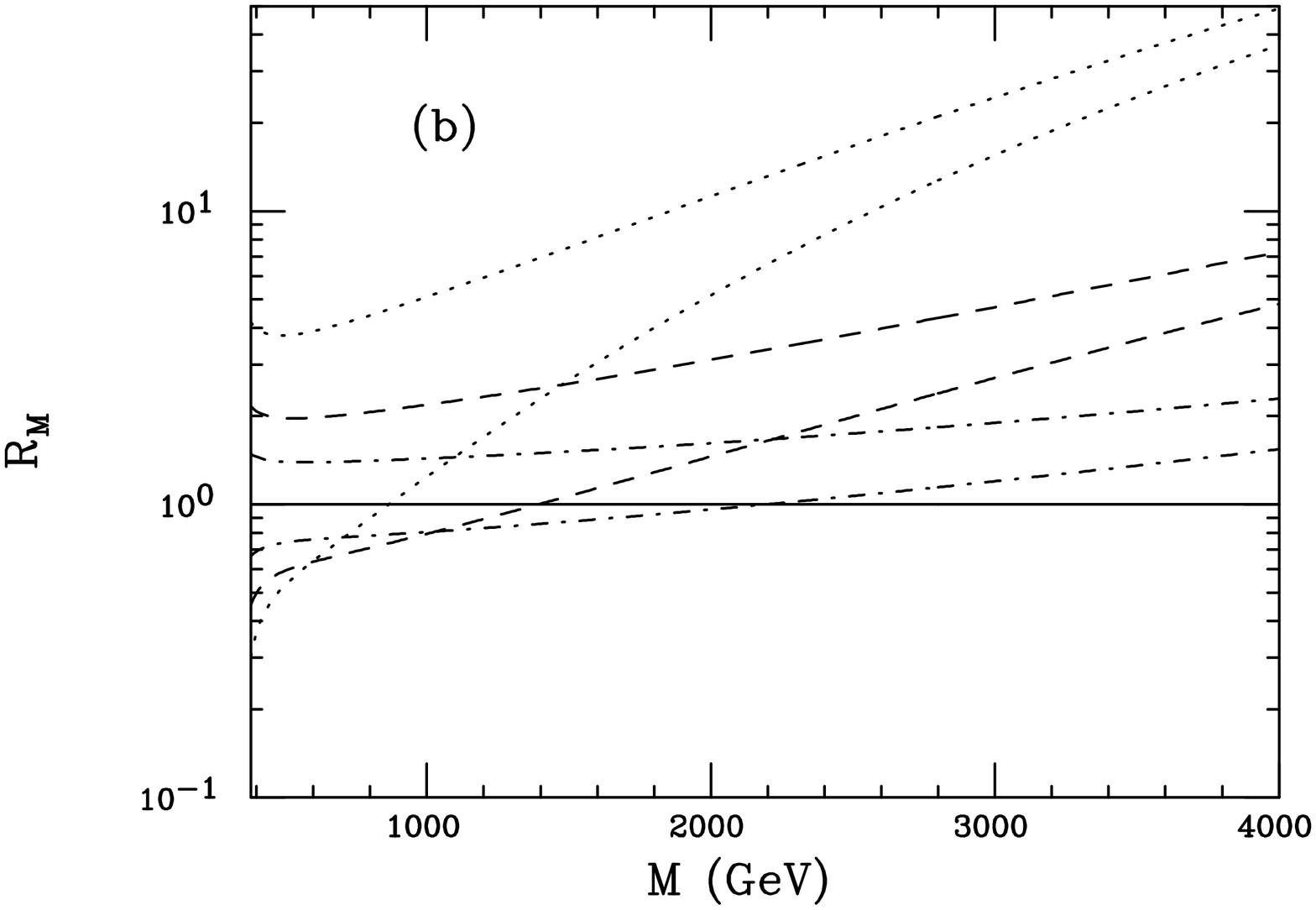,height=9.1cm,width=9.1cm,angle=-90}}
\vspace*{0.1cm}
\centerline{
\psfig{figure=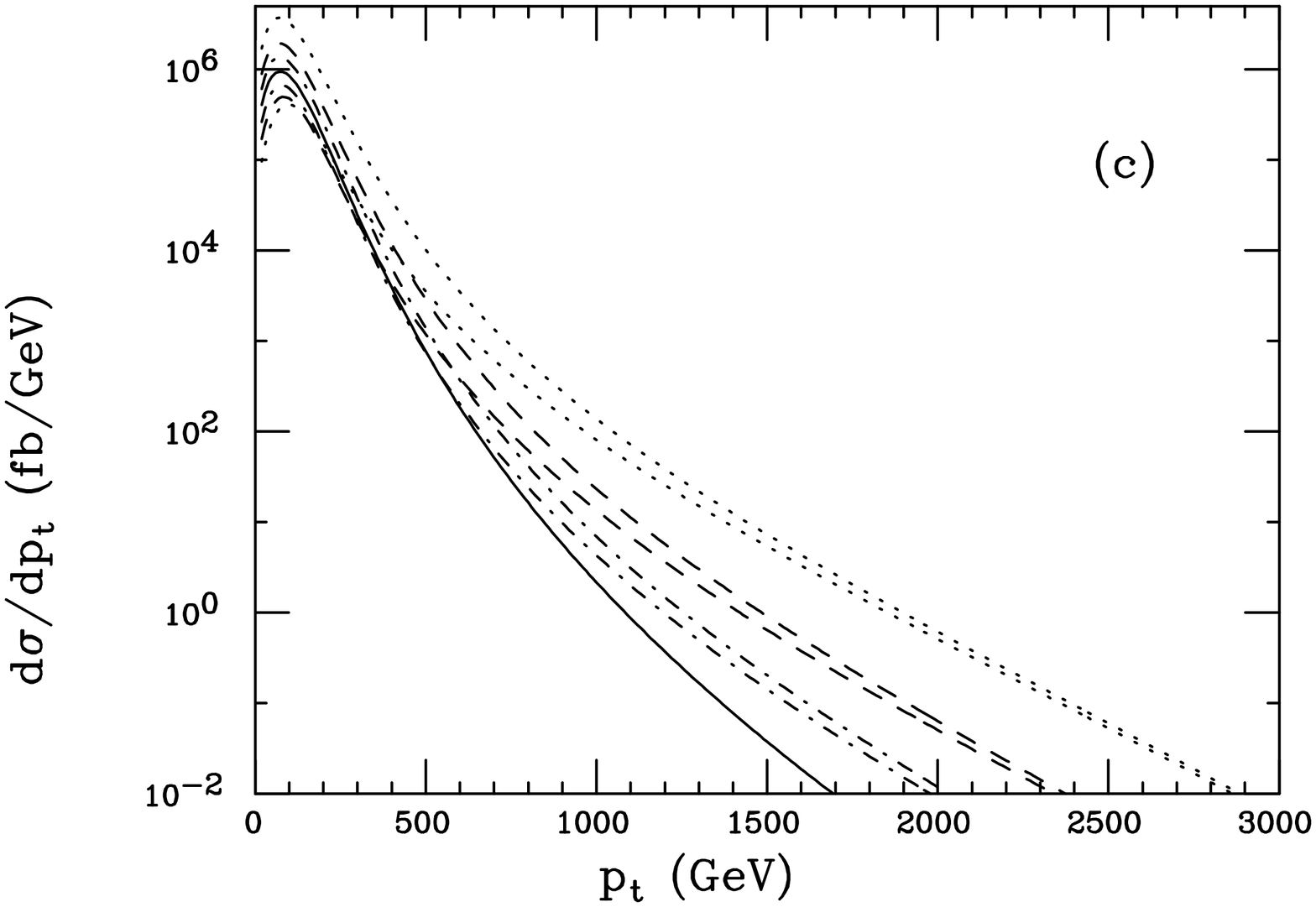,height=9.1cm,width=9.1cm,angle=-90}
\hspace*{-5mm}
\psfig{figure=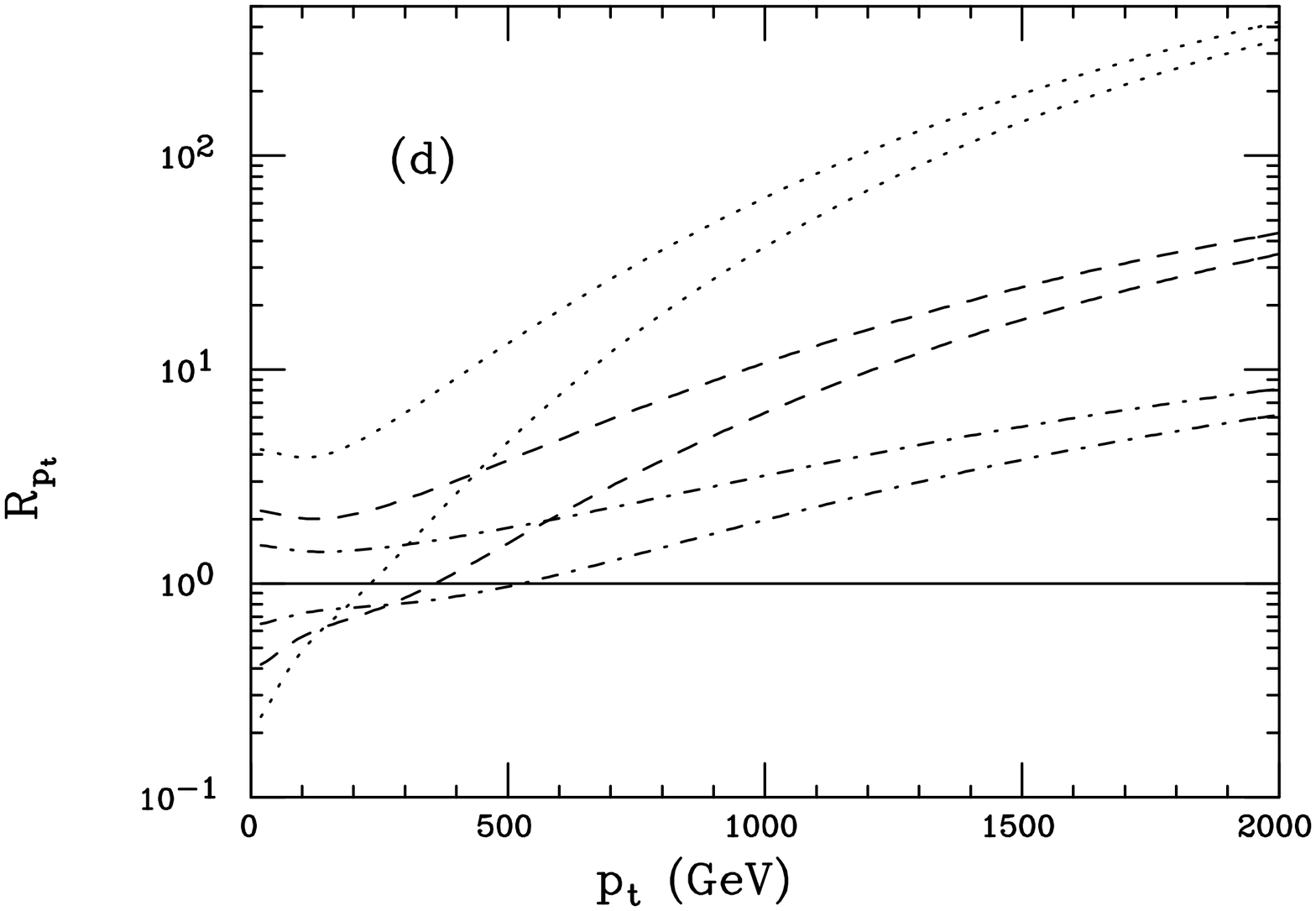,height=9.1cm,width=9.1cm,angle=-90}}
\vspace*{-0.5cm}
\caption{\small (a) $t\bar t$ invariant mass distribution at the LHC for 
various values of $\kappa$ assuming $m_t=180$ GeV. (b) The same distribution 
scaled to the SM result. (c) $t\bar t$ $p_t$ distribution at the LHC and (d) 
the same distribution scaled to the SM. In all cases, the SM is represented 
by the solid curve whereas the upper(lower) pairs of dotted(dashed, 
dash-dotted) curves  
corresponds to $\kappa=$0.5(-0.5), 0.25(-0.25), and 0.125(~-0.125), 
respectively.}
\label{dislhc}
\end{figure}
\vspace*{0.4mm}

\section{Analysis}

As has been shown elsewhere{\cite {big}}, the $p_t$ and pair invariant 
mass ($M_{tt}$) distributions for top quark pair production at the LHC are 
highly sensitive to non-zero values of $\kappa$. Figures~\ref{dislhc}a 
and ~\ref{dislhc}c show the 
modifications in the SM expectations for both $d\sigma/dM_{tt}$ and 
$d\sigma/dp_t$, respectively, for different values of $\kappa$. 
Perhaps more revealingly, Figures~\ref{dislhc}b and~\ref{dislhc}d show the 
ratio of the modified distributions to the corresponding SM ones. We see the 
important results that a non-zero $\kappa$ leads to ($i$) enhanced cross 
sections at large $p_t$ and $M_{tt}$ and ($ii$) the {\it shapes} of the 
distributions are altered, \ie, the effect is not just an overall change in 
normalization. This is contrary to what was observed in the Tevatron case 
where both $d\sigma/dM_{tt}$ and $d\sigma/dp_t$ were essentially just rescaled 
by the ratio of the total cross sections. Clearly, data on these two 
distributions at the LHC can lead to significant constraints on $\kappa$ or 
observe a non-zero effect if $\kappa$ is sufficiently large. In 
Ref.~{\cite {big}}, the $cos \theta^*$ and rapidity($\eta$) distributions 
were also examined but they were found to be less sensitive to non-zero 
$\kappa$ than the dramatic effects shown in Fig.~\ref{dislhc}. 

\vspace*{-0.5cm}
\nn
\begin{figure}[htbp]
\centerline{
\psfig{figure=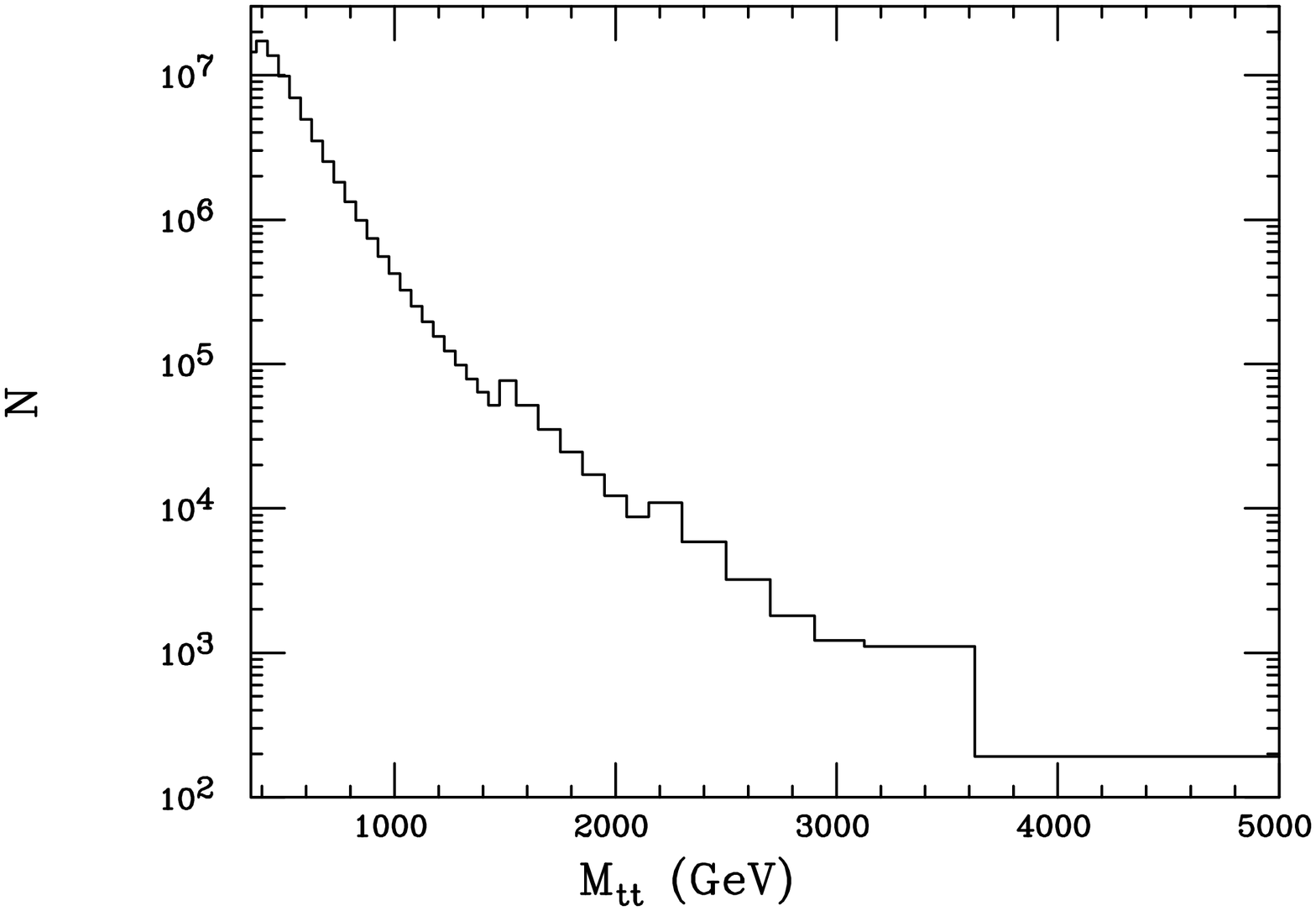,height=9.1cm,width=9.1cm,angle=-90}
\hspace*{-5mm}
\psfig{figure=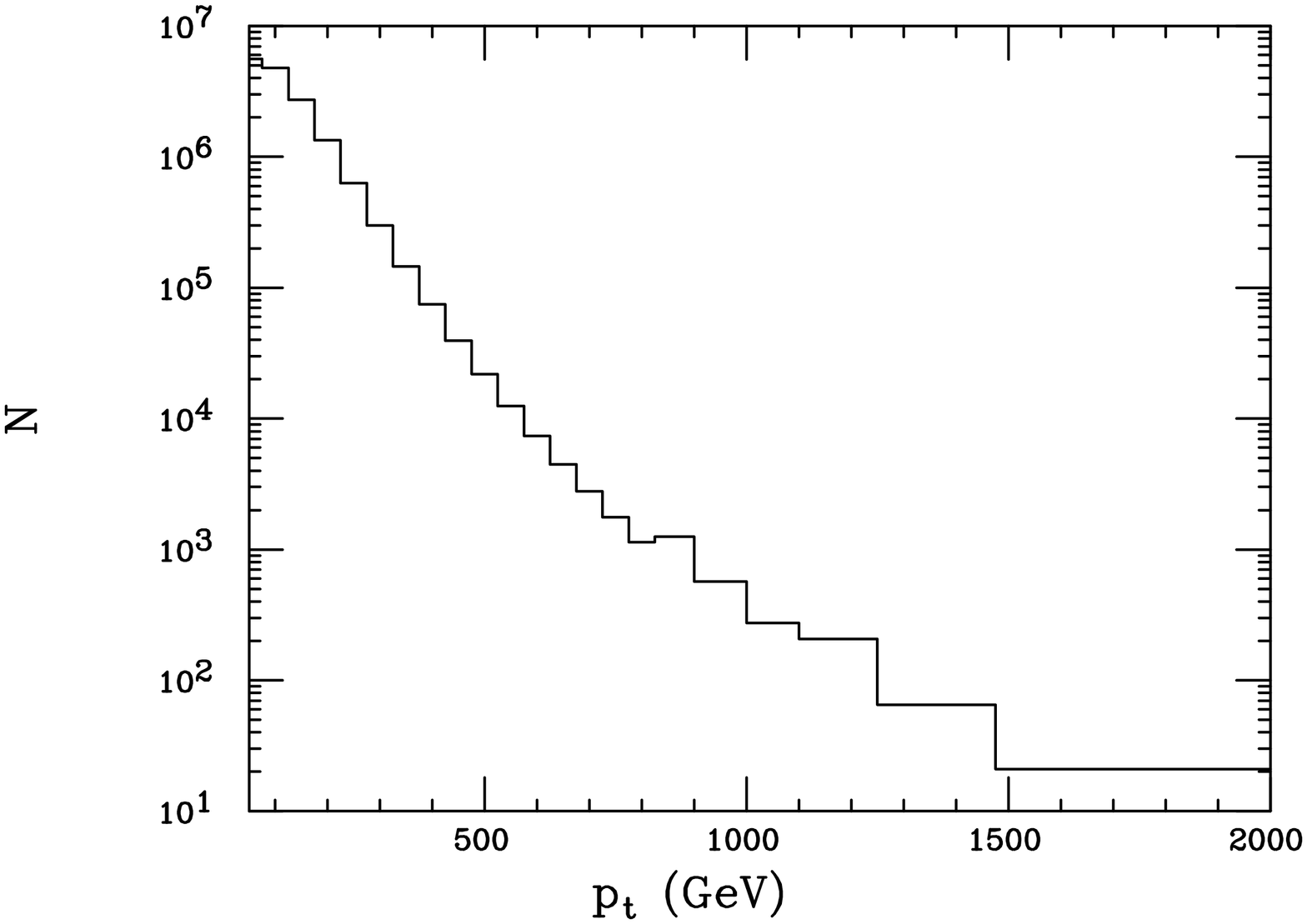,height=9.1cm,width=9.1cm,angle=-90}}
\vspace*{-0.1cm}
\caption{Sample histograms of top quark data generated for the LHC, assuming 
100 $fb^{-1}$ of integrated luminosity. On the left(right) is the top pair 
invariant mass ($p_t$) distribution. MRSA$'$ parton densities and $m_t=175$ 
GeV have been assumed.}
\label{histlhc}
\end{figure}
\vspace*{0.4mm}

How sensitive are these distributions to non-zero $\kappa$ and what bounds can 
be obtained at the LHC? 
In order to answer these questions, we follow a Monte Carlo approach. We begin  
by generating 100 $fb^{-1}$ `data' samples for both distributions 
{\it assuming} the SM is correct. To be specific, since the next to leading 
order(NLO) expressions for these distributions in the presence of anomalous 
couplings do not yet exist, we use the leading order results rescaled by the 
NLO/LO cross section ratios for both subprocesses as effective $K$-factors to 
obtain a rough estimate of these higher order effects.
Sample histograms of this appropriately rescaled `data' are shown 
in Fig.~\ref{histlhc}. Note that 
there are 37 bins in $M_{tt}$ and 22 bins in $p_t$ of varying sizes 
essentially covering the entire kinematically allowed ranges. Bin sizes are 
adjusted to partially conform to changes in resolution and declining 
statistics as we go to larger values of either kinematic variable. In addition 
to the usual statistical errors, we attempted to include some estimate of the 
systematic point-to-point errors. These were added in quadrature to the 
statistical errors. Thus, 
neglecting the overall normalization uncertainties, the 
error in the number of events($N_i$) in a given bin($i$) was assumed to be 
given by 
\begin{equation}
\delta N_i= [N_i+aN_i^2]^{1/2}
\end{equation}
with the parameter $a$ setting the {\it a priori} unknown size of the 
systematic error. Note that we 
have made the simplifying assumption that the magnitude of $a$ is bin 
independent. The total 
error is thus generally systematics dominated. With these errors the Monte 
Carlo generated data 
was then fit to the known functional form of the relevant distribution: 
\begin{equation}
{d\sigma \over {d\cal O}}= f\sum_{n=0}^{4} \kappa^n g_n(\cal O)
\end{equation}
where $f$ allows the overall normalization to float in the fit and the $g_n$ 
were those appropriate to either the $p_t$ or $M_{tt}$ distributions. 

\begin{figure}[htbp]
\centerline{
\psfig{figure=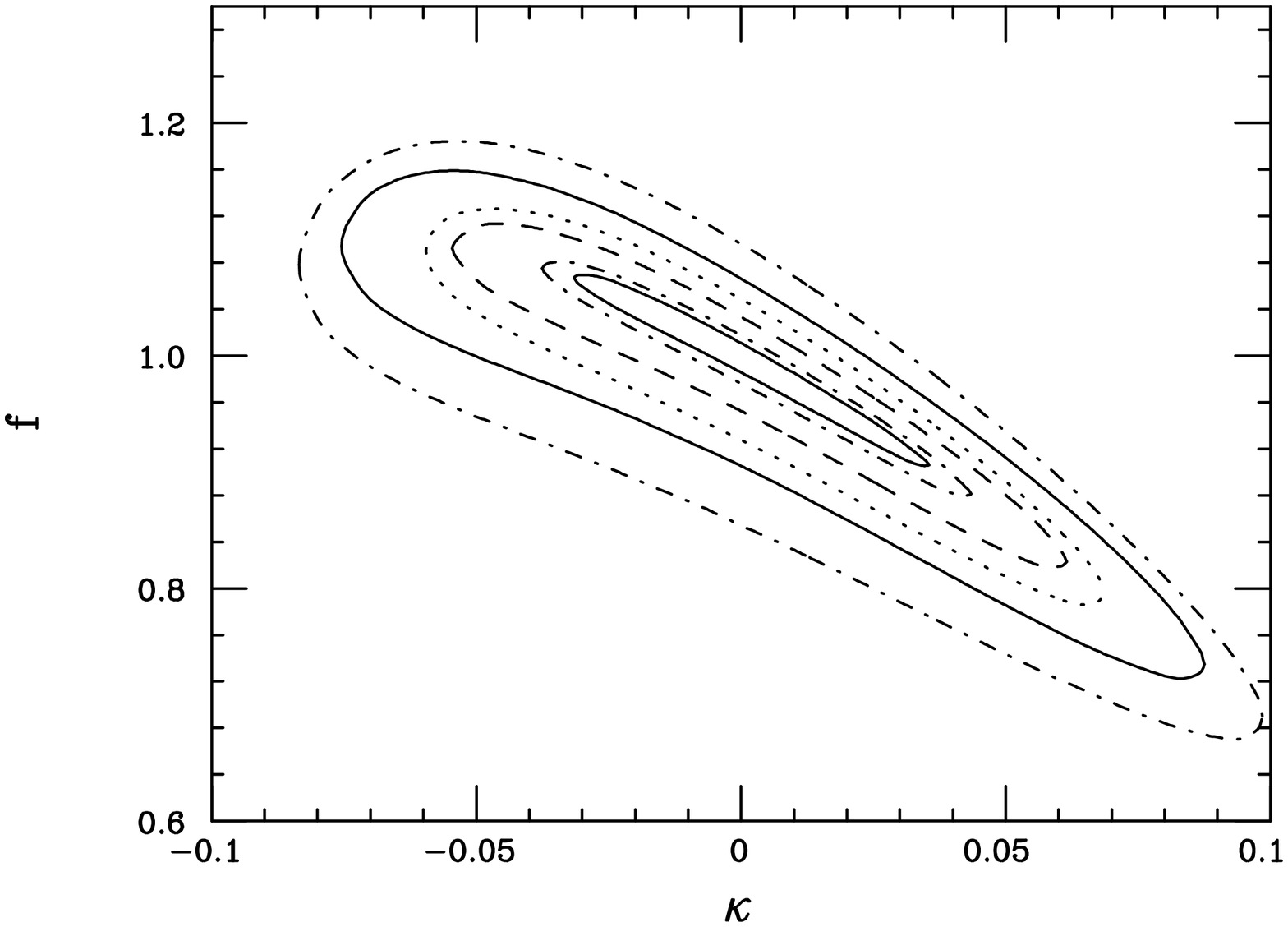,height=9.1cm,width=9.1cm,angle=-90}
\hspace*{-5mm}
\psfig{figure=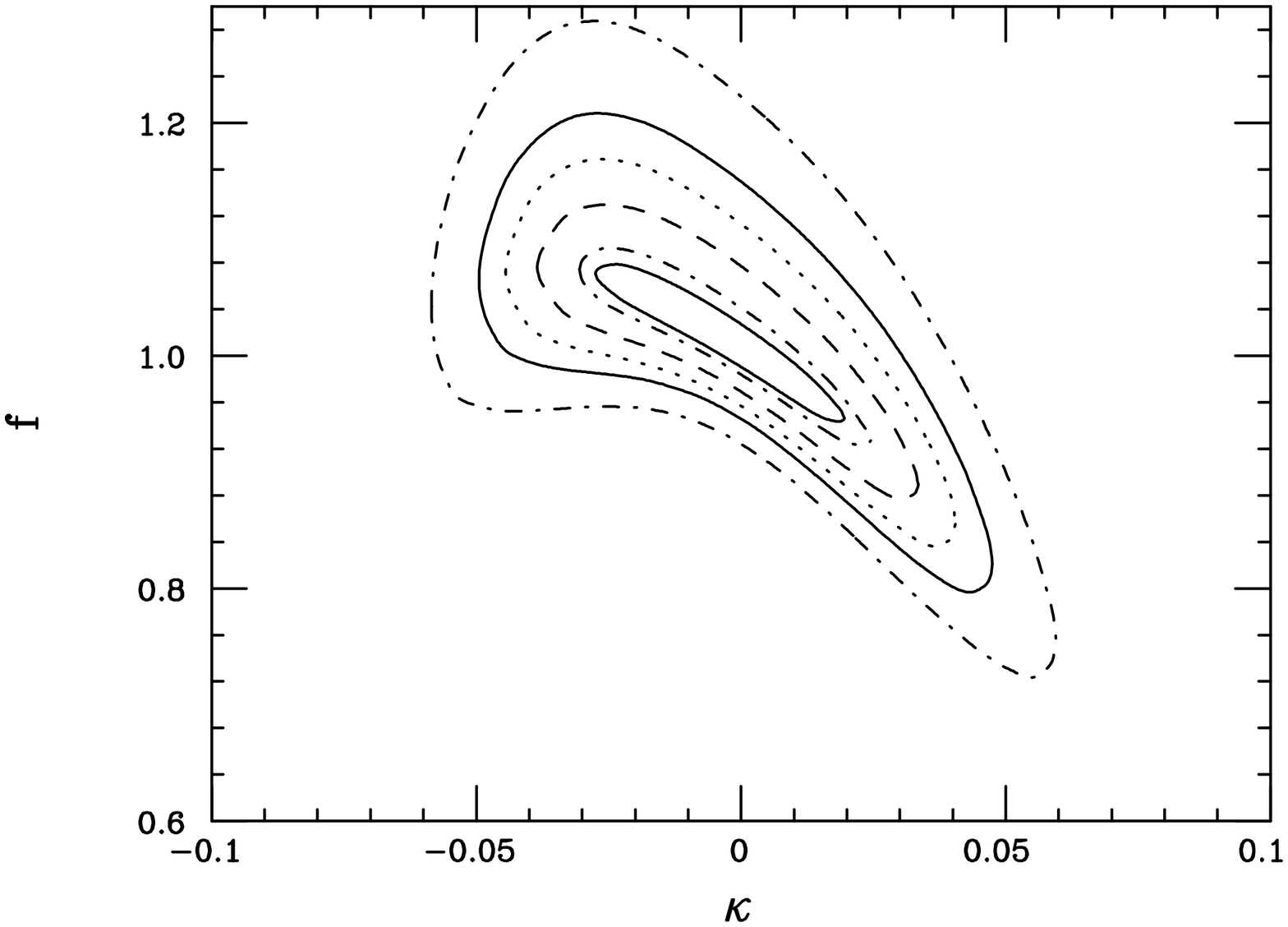,height=9.1cm,width=9.1cm,angle=-90}}
\vspace*{-0.1cm}
\caption{$95\%$ CL two parameter $(f,\kappa)$ fits to the invariant mass(left) 
and $p_t$(right) distributions at the LHC for a 175 GeV top quark assuming the 
MRSA$'$ parton densities for different assumed values of the systematic errors 
parameterized by $a$. From inside out the curves correspond to $a$= 0.03, 
0.05, 0.10, 0.15, 0.20, and 0.30, respectively. }
\label{reslhc}
\end{figure}

The results of this analysis are thus a set of $95\%$ CL allowed regions in the 
$f-\kappa$ plane for 
various assumed values of the anticipated size of the systematic errors. These 
can be seen in Figure~\ref{reslhc}. Here we see that for systematic errors of 
reasonable magnitude the value of $\kappa$ is constrained to lie in the range 
$-0.09 \leq \kappa \leq 0.10$ from the $M_{tt}$ distribution and  
$-0.06 \leq \kappa \leq 0.06$ from the corresponding $p_t$ distribution. Note 
that the correlation between $f$ and $\kappa$ is much stronger in the case of 
the $M_{tt}$ distribution. Increasing the integrated luminosity by a factor of 
two will not greatly affect our results since the errors are systematics 
dominated.  Combining the results of multiple distributions in a global fit to 
$\kappa$ will most likely result in even strong bounds. 

\vspace*{-0.5cm}
\nn
\vspace*{0.4mm}
\vspace*{-0.5cm}
\nn
\begin{figure}[htbp]
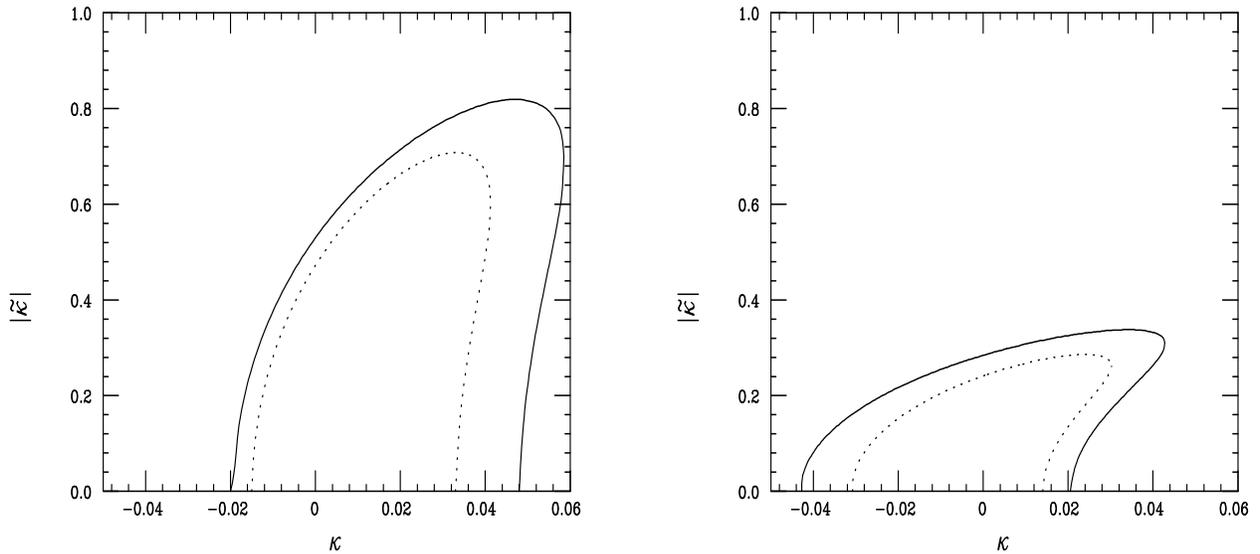

\centerline{
\psfig{figure=ttnewg.res1ps,height=9.1cm,width=9.1cm,angle=-90}
\hspace*{-5mm}
\psfig{figure=ttnewg.res2ps,height=9.1cm,width=9.1cm,angle=-90}}
\vspace*{-0.1cm}
\caption{$95\%$ CL allowed regions in the $\kappa-\tilde \kappa$ plane 
obtained from fitting the gluon spectrum. On the left the fit is for 
gluon jets above $E_g^{min}$=25 GeV at a 
500 GeV NLC assuming an integrated luminosity of 50(solid) or 100(dotted) 
$fb^{-1}$. On the right is the case of a 1 TeV collider with $E_g^{min}$=50 
GeV and luminosities of 100(solid) and 200(dotted) $fb^{-1}$. Note that the 
allowed region has been significantly compressed downward in comparison to 
the 500 GeV case.}
\label{resnlc}
\end{figure}
\vspace*{0.4mm}

Using Figure~\ref{resnlc} we can make a direct comparison of the bounds 
obtainable on $\kappa$ at the NLC by using the process 
$e^+e^- \to t\bar tg$ as discussed in Ref.{\cite {tgr}} with those from the 
LHC analysis above. In these Figures the influence of $\tilde \kappa$ is 
also shown. These NLC results were obtained by fitting the 
spectrum of very high 
energy gluon jets produced in association with top pairs (above some cut,  
$E_g^{min}$, used to avoid contamination from the radiation off final state 
$b$-quarks in top decay). Only statistical errors were included in the 
analysis. The resulting bounds are essentially statistics limited. We see 
from these Figures that the constraints on $\kappa$ from the $\sqrt s$=500 GeV 
NLC with an integrated luminosity of 50 $fb^{-1}$ are only slightly better 
than what is achievable at the LHC from the top pair's $p_t$ distribution. 
The constraints tighten at the 1 TeV NLC. Clearly the LHC and NLC have 
comparable sensitivities to the anomalous chromomagnetic moment of the top.

\section{Acknowledgements}

The author would like to thank J. Hewett, A. Kagan, P. Burrows, L. Orr, and 
R. Harris for discussions related to this work.

%
\def\MPL #1 #2 #3 {Mod.~Phys.~Lett.~{\bf#1},\ #2 (#3)}
\def\NPB #1 #2 #3 {Nucl.~Phys.~{\bf#1},\ #2 (#3)}
\def\PLB #1 #2 #3 {Phys.~Lett.~{\bf#1},\ #2 (#3)}
\def\PR #1 #2 #3 {Phys.~Rep.~{\bf#1},\ #2 (#3)}
\def\PRD #1 #2 #3 {Phys.~Rev.~{\bf#1},\ #2 (#3)}
\def\PRL #1 #2 #3 {Phys.~Rev.~Lett.~{\bf#1},\ #2 (#3)}
\def\RMP #1 #2 #3 {Rev.~Mod.~Phys.~{\bf#1},\ #2 (#3)}
\def\ZP #1 #2 #3 {Z.~Phys.~{\bf#1},\ #2 (#3)}
\def\IJMP #1 #2 #3 {Int.~J.~Mod.~Phys.~{\bf#1},\ #2 (#3)}

\end{document}